\documentclass[12pt]{iopart}

\begin{document}

\title{Entropic theory of Gravitation}

\author{Victor Atanasov}

\address{Department of Condensed Matter Physics, Sofia University, 5 boul. J. Bourchier, 1164 Sofia, Bulgaria}
\ead{vatanaso@phys.uni-sofia.bg}



\begin{abstract}
We construct a manifestly Machian theory of gravitation on the foundation that information in the universe cannot be destroyed (Landauer's principle). If no bit of information in the Universe is lost, than the sum of the entropies of the geometric and the matter fields should be conserved. We propose a local invariant expression for the entropy of the geometric field and formulate a variational principle on the entropic functional which produces entropic field equations. This information-theoretic approach implies that the geometric field does not exist in an empty of matter Universe, the material entropy is geometry dependent, matter can exchange information (entropy) with the geometric field and a  quantum condensate can channel energy into the geometric field at a particular coherent state. The entropic field equations feature a non-intuitive direct coupling between the material fields and the geometric field, which acts as an entropy reservoir. Cosmological consequences such as the emergence of the cosmological constant as well as experimental consequences involving gravity-quantum condensate interaction are discussed. The energetic aspect of the theory restores the repertoire of the classical General Relativity up to a different coupling constant between the fields.
\end{abstract}


\section{Motivation}

Information is physical, that is the erasure of one bit
of information increases the
entropy of the Universe by $k_B \log2,$ where $k_B$ is Boltzmann's constant, that is the Landauer's principle\cite{Landauer}. By bit here we will understand a unit of information, a variable (or binary choice) that can take only the values 0 or 1 and
by information content of an object - the size of the set of
instructions (number of bits) necessary to construct the object or its state.
The Landauer's principle was confirmed experimentally\cite{exp_L} and led to the resolution of the Maxwell's demon paradox\cite{Bennett}, in our view an example of not well explored connection between physics and information.

Maxwell's demon collects information about a system, lowers its entropy and increases its energy. The demon's operation has a  net result of total conversion of heat into work, a process forbidden by the second law of thermodynamics. Bennett showed, that measurements performed by the demon in order to prepare the system for energy extraction can be done without  doing work but that leaves the only one other possibility consistent with the second law, namely the erasure step, required to return the demon's memory (erase his knowledge) to its original state necessary to perform the cycle again, is dissipative. Thus, Landauer's principle resolves a century old thermodynamic gedanken paradox: the work gained by the engine $W_{extracted} = k_B T \log 2$ is converted into heat in the process of erasure of information from the demon’s physical mind $W_{erasure} = - k_B T \log 2$, therefore no net work is
produced in the cycle.

The resolution of the Maxwell's demon paradox with the help of the Landauer's principle points to the equivalence of conservation of energy and the conservation of information. However, the conservation of information 
\begin{eqnarray}
\sum_j^{N} I_j=\textrm{const.},
\end{eqnarray}
where $I_j$ is the information associated with a physical object or field, is not an explored venue in physics and its consequences are largely unknown. In this paper we will suppose that the information content of the Universe is constant and employ a variational principle to find the entropic equations of motion as information lost in one system is transferred as information gained in other system
\begin{eqnarray}
\sum_j^{N} \delta I_j= - \sum_j^{N} \delta S_j=0.
\end{eqnarray}
These systems might be a quantum condensate and a geometric, that is gravitational field. We will introduce a concept of information content (entropy) of the geometric field. The concept of information content of the matter fields is given by the von Neumann entropy. Throughout this paper by information we will understand
\begin{eqnarray}
I=S_{\textrm{max} } - S,
\end{eqnarray}
that is the difference between the maximal possible entropy of the system (total disorder; no knowledge of the microstate the system is in) and the temporary entropy. 

The introduction of an appropriate expression for the entropy of the geometric field goes through the recognition that an entropy can be associated with a black hole. This entropy is
equal to a quarter of the horizon area in Planck units of length\cite{Bekenstein}:
\begin{eqnarray}
S \propto \textrm{Area of the Horizon}
\end{eqnarray}
It has also been noted that the dependence of gravitational entropy on area is not
restricted to black holes only but can be asociated with the entanglement entropy in quantum (conformal) field theories via AdS/CFT correspondence\cite{Ryu}. 

Assuming a number of restrictions, the entropy in a region of space is limited
by the area of its boundary, that is the holographic principle\cite{holographic}. In other words, the entropy bound\cite{Bekenstein} restricts 
that the physics in a region of space is characterised by the amount of information that fits on its boundary surface, at one bit per unit of Planck area $l_p^2 \approx 2.6 \times 10^{-66}$ cm$^2.$ Susskind's use of the term "hologram" is justified by the hypothetic existence of preferred surface  in space-time, kind of cosmological horizon, onto which all of the data in the universe can be imprinted at the same rate of a bit per unit of Planck area.

Alternatively, the main ingredient necessary to derive gravity is information\cite{Verlinde}.  In Verlinde's view that is the amount of information associated with matter and its location, measured in terms of entropy, defined as the holographic principle dictates. Changes in the entropy when matter is displaced, that is the change in the interference pattern of the hologram on the cosmological screen, leads to an entropic force, which takes the form of gravity. As the author himself admits, to a degree the result is tautological, that is traveling the path backwards from the final destination of black hole thermodynamics and the holographic principle, one is assured to end up at the beginning - the laws of gravity. However, the notion that gravity might actually be an entropic force is compelling.

Unfortunately, the holographic principle is not an idea that might lead to a local theory. Experimentally verified theories in physics
are local (excluding entanglement). In local theories the number
of degrees of freedom is proportional to volume. The holographic principle postulates that the entropic content is proportional to 
the area of surface enclosing the embedding volume\cite{Bousso}. 

We believe, that this weakness is mainly a result of the lack of an appropriate local covariant expression for the entropy of space-time. In the present article we propose such an expression and explore some of its consequences. We retain Verlinde's proposition that entropy might actually be an entropic force but re-write its formulation. We also re-write the standard approach to deriving gravity: instead of formulating a variational principle based on energy conservation, we formulate a variational principle based on information conservation.

\section{Theory}

Let us assume that the entropy of the geometric field is given by the expression
\begin{eqnarray}\label{eq:S_G}
S_G=\frac{1}{L^2 R} 
\end{eqnarray}
which is manifestly covariant, provided $R$ is the Ricci scalar curvature and $L$ is a unit of length, which at the present stage is a parameter in the theory to be built. This parameter might be equal to the Plank length or the Einstein length $L=\sqrt{1/\Lambda},$ where $\Lambda \approx 1.19 \times 10^{-52} \textrm{m}^2$ is the cosmological constant, proportional to the energy density of the vacuum.  The inverse of the Ricci scalar curvature is a local geometric object with the dimension of surface area
\begin{eqnarray}
\frac{1}{R}  \propto \textrm{Area} .
\end{eqnarray}
Indeed, this form of the local entropy suffers from a divergence in the case of $R=0,$ that is flat space is associated with maximal infinite entropy, but this situation can be remedied by offsetting the singular case
\begin{eqnarray}
S_G=\frac{1}{L^2 \left(R + R_0 \right) } .
\end{eqnarray}
This fixes the entropy of flat space to some maximal constant value. We are not going to explore such  theory with a cut-off and focus on the main consequences of (\ref{eq:S_G}). 

The expression for the geometric entropy can serve as a density to a more generalised estimate of the geometric entropy
\begin{eqnarray}
&& \mathcal{S_G}=\int S_G  \sqrt{-g} d^4 x
\end{eqnarray}
which will be the main ingredient of our theory
\begin{eqnarray}\label{eq:theory}
&& \int S_G \sqrt{-g} d^4 x+ \int S_{QM} \sqrt{-g} d^4 x =\mathrm{const},
\end{eqnarray}
where $S_{QM}$ is the entropy density of the material fields. This is the mathematical expression of the idea that information in the universe cannot be destroyed but transferred from the quantum system to the entropy reservoir, that is the geometric field, and back.

The variation of the geometric entropy amounts to
\begin{eqnarray}
\nonumber \delta \sqrt{-g} S_G&=&\frac{f'(R)}{L^2}  \sqrt{-g}  \delta \left(  g_{\mu \nu} \right) R^{\mu \nu} +\frac{f(R)}{L^2}    \delta \sqrt{-g} + \frac{1}{L^2}  \left( f'(R)    \sqrt{-g} g_{\mu \nu} \right) \delta  R^{\mu \nu},
\end{eqnarray}
where $f'(R)=-1/R^2$ is the derivative with respect to $R$ of $f(R)=1/R.$ This expression can be varied with respect to the components of the metric tensor 
and the coefficients of the affine connection upon recognition of the relation $\delta R_{\mu \nu}=-\nabla_{\alpha} \delta \Gamma_{\mu \nu}^{\alpha} + \nabla_{\nu} \delta \Gamma_{\mu \alpha}^{\alpha} $ and integration by parts\cite{Schrodinger}
\begin{eqnarray}\label{eq: delta S_G / delta Gamma} 
\frac{\delta \sqrt{-g}  S_G}{\delta \Gamma^{\alpha}_{\mu \nu}}&=& \frac{1}{L^2} \left[ \nabla_{\alpha}\left( f'(R)\sqrt{-g} g^{\mu \nu} \right) \right. - \frac12 \nabla_{\nu}\left( f'(R) \sqrt{-g} g^{ \mu \nu} \right) \delta^{\nu}_{\alpha}  \\
\nonumber  && \left. - \frac12 \nabla_{\mu}\left( f'(R) \sqrt{-g} g^{ \nu \mu} \right) \delta^{\mu}_{\alpha}
  \right],
\end{eqnarray}
Here we have symmetrised the above relation with respect to the indices $\mu,\nu.$ We have used the Palatini variational method as a mathematical tool to treat the theory at hand and by introducing new independent fields (the coefficients of affine connection) to lower the order of the entropic field equations.

Its variation with respect to the metric is given by
\begin{eqnarray}\label{eq: delta S_g / delta g}
\nonumber \frac{\delta \mathcal{S_G}}{\delta g_{\mu \nu}} &=& \frac{1}{L^2} \left[ f'(R) R^{\mu \nu} - \frac12 g^{\mu \nu} f(R)  \right]=-\frac{1}{L^2 R^2} \left[ R^{\mu \nu} + \frac12 g^{\mu \nu} R  \right]=0.
\end{eqnarray}
Note, setting  it to zero means that the information we have for the space-time is maximal, that is the entropy is minimal and therefore its change (variation) as the embedding space changes its lengthscale is also vanishing. As is well known, for the singularities of space-time $R \to \infty$ the ''no hair theorem'' applies which confirms the result here, namely singularities in space-time are actually simpler and characterised by a minimal set of information\cite{Israel}. We only need to know  three degrees of freedom, namely mass, electric charge and angular momentum.

Indeed the ''vacuum'' entropic equation of motion
\begin{eqnarray}\label{eq: S_G vacuum}
\lim_{R \to \infty}{S_G}=0   \qquad  & \& & \qquad  \lim_{ R\to \infty} \frac{\delta \mathcal{S_G}}{\delta g_{\mu \nu}} = 0
\end{eqnarray}
is vanishing when the geometry of space-time is singular, that is in this entropic theory of gravity $R=0$ corresponds to infinite entropy of space-time and infinite variation. 
\begin{eqnarray}\label{eq: S_G matter}
\lim_{R \to 0}{S_G}=\infty   \qquad  & \& & \qquad \lim_{ R\to 0} \frac{\delta \mathcal{S_G}}{\delta g_{\mu \nu}} = \infty
\end{eqnarray}
In effect, flat space-time appears to be an entropic reservoir with infinite capacity and the singular points are regions where the entropy and its variation are vanishing in accord with the classical picture. Disorder is vanishing and the information about space-time is maximal at singularities. However,  $R=0$ and $R=\infty$ do not define affine connection. In order for affine connection to emerge in the flat space-time case matter content is necessary. The same is true for the singularity case.

The variation with respect to the affine connection coefficients yields\cite{Vollick, Schrodinger}
\begin{eqnarray}\label{eq:vollick}
\frac{\delta \sqrt{-g}  S_G}{\delta \Gamma^{\alpha}_{\mu \nu}}=0 \qquad  &\Rightarrow&  \qquad  
\nabla_{\alpha}\left( f'(R)\sqrt{-g} g^{\mu \nu} \right)=0,
\end{eqnarray}
by multiplying with $g^{\nu \alpha}$ and contracting the indices $\nu, \alpha$ with the help of the Kronecker symbol. This relation can be identified as the equivalent of $\nabla_{\alpha} \tilde{g}^{\mu \nu} =0$ and entails
\begin{eqnarray}
\Gamma^{\alpha}_{\mu \nu} = \left\{\begin{array}{c} \alpha \\ \mu \;  \nu \end{array}\right\}_{ \textrm{with respect to} \;  \tilde{g} },
\end{eqnarray}
that is the $\Gamma$'s are given by the Christoffel brackets provided there exists a new metric $\tilde{g}^{\mu \nu}=f'(R)\sqrt{-g} g^{\mu \nu}.$  Transforming conformally the metric $\tilde{g}_{\mu \nu}= g_{\mu \nu}/(f'(R)\sqrt{-g}) $ back to the metric $g_{\mu \nu}$ leads to\cite{Hawking}
\begin{eqnarray}\label{eq:connection}
 \Gamma^{\alpha}_{\mu \nu}= \left\{\begin{array}{c} \alpha \\ \mu \;  \nu \end{array}\right\} -   \left[  \delta^{\alpha}_{(\mu} \partial_{\nu )} - \frac{1}{2} g_{\mu \nu} g^{\alpha \sigma} \partial_{\sigma} \right]\log\left( f'(R) \sqrt{-g} \right) &&
\end{eqnarray}
Here we have symmetrised with respect to the indices $\mu, \nu$ and the first term is the Christoffel bracket with respect to the metric $g_{\mu \nu}.$

Since the vacuum equations 
\begin{eqnarray}
\frac{\delta \sqrt{-g}  S_G}{\delta \Gamma^{\alpha}_{\mu \nu}}=0: \qquad  \left.S_G\right|_{R \to 0}=\infty  \quad & \& & \quad  \not{\exists} \; \Gamma^{\alpha}_{\mu \nu}
\end{eqnarray}
imply that for $R\to \infty$  the expression $\partial_{\alpha} \log{f'(R)} \to \infty,$ as well as
$R\to 0$ leads to $\partial_{\alpha} \log{f'(R)} \to 0$ we have no solution for 
$\Gamma^{\alpha}_{\mu \nu}.$

An important point that needs to be addressed is that (\ref{eq:connection}) does not appear to define the connection since $f'(R)$ as a function of the Ricci scalar  $R$ involves the derivatives of the connection. Therefore, in any theory we should be able to express $R$ in terms of the quantum matter variation of entropy which should contain only components and derivatives of the metric $g_{\mu \nu}.$ Note, the vacuum case implies that thermodynamically preferred is the state of non-existence of the geometric field. A  space-time can emerge only in the presence of matter
\begin{eqnarray}\label{eq: delta Theory / delta Gamma}
\frac{\delta \sqrt{-g}  S_G}{\delta \Gamma^{\alpha}_{\mu \nu}}=-\frac{\delta \sqrt{-g}  S_{QM}}{\delta \Gamma^{\alpha}_{\mu \nu}}
\end{eqnarray}
which is both an equation for the coefficients of the affine connection and an indication that the theory is manifestly in accord with the ideas of Ernst Mach and bishop G. Berkeley, namely the local geometry is determined by the matter distribution in the Universe. According to Bondi-Samuel classification of statements of the Mach principle, Mach 7 fits the ideas here very closely \cite{BS}: {\it If one takes away all matter, there is no more space.} Note, by lack of space our theory assumes $\not{\exists} \; \Gamma^{\alpha}_{\mu \nu}.$ Therefore, the divergence of the expression for the geometric entropy for $R \to 0$ is not a serious drawback of the theory. We will show that {\it the presence of matter will introduce a cut-off which is intimately associated with the cosmological constant.}

Now let us explore the case when quantum matter, which carries its amount of information (entropy) fills space and can exchange information (entropy) with the entropy reservoir, namely the geometric field.

Looking for the stationary points of the information functional (\ref{eq:theory}) by varying with respect to the metric components $g_{\mu \nu}$ yields
\begin{eqnarray}
\frac{1}{L^2 R^2} \left[ R^{\mu \nu} + \frac12 g^{\mu \nu} R  \right] = - \frac12 g^{\mu \nu} S_{QM}  +  \frac{\delta S_{QM}}{\delta g_{\mu \nu}}
\end{eqnarray}
Contracting the indices by multiplying with $g_{\mu \nu}$ produces a scalar theory
\begin{eqnarray}\label{eq: R}
R= \frac{3}{ L^2 \left[ - 2 S_{QM}  +  \frac{\delta S_{QM}}{\delta g} \right] }
\end{eqnarray}
which gives an expression for
\begin{eqnarray}
f'(R)=-\frac{1}{R^2}= - L^4 \left[ - \frac23 S_{QM}  + \frac13 \frac{\delta S_{QM}}{\delta g} \right]^2 
\end{eqnarray}
in terms of the matter content. This expression for $f'(R)$ is a function of the metric and its derivatives at most, therefore (\ref{eq:connection}) defines affine connection and completes the geometric field. 

However, the affine connection is governed by (\ref{eq: delta Theory / delta Gamma}). The geometric field part is given by (\ref{eq: delta S_G / delta Gamma}), now we need to find out an expression for the material part. Let us consider the simplest case of a non-interacting quantum condensate which possesses only kinetic energy 
\begin{eqnarray}\label{eq:LB g&Gamma}
H= - \frac{\hbar^2}{2m} \Delta_{LB}= - \frac{\hbar^2}{2m} \left (  \hat{g}^{jk}  \partial_{j} \partial_{k} - \hat{g}^{jk}  \hat{\Gamma}^{l}_{jk} \partial_{l}   \right) , 
\end{eqnarray}
where $\Delta_{LB}$ is the Laplace-Beltrami operator defined on a three-dimensional hyperplane of space-time. $\hat{g}_{jk}$ is the three-dimensional metric and $\hat{\Gamma}^{l}_{jk}$ is the Christoffel symbol defined by $\hat{g}$. Provided there is curvature in the embedding space the kinetic energy of the quantum matter becomes length-scale dependent. 

The variation of the energy operator with respect to the metric components is given by
$\delta H = - \frac{\hbar^2}{2m} \delta \hat{g}^{jk}  \left (    \partial_{j} \partial_{k} -   \hat{\Gamma}^{l}_{jk} \partial_{l}   \right) 
$ which yields
\begin{eqnarray}
\frac{\delta H}{\delta g^{\mu \nu }} = - \frac{\hbar^2}{2m}  \left (    \partial_{\mu} \partial_{\nu} -  \hat{ \Gamma}^{l}_{\mu \nu} \partial_{l}   \right) 
\end{eqnarray}
Note the variation of the three dimensional metric with respect to the four dimensional one is easily defined on 3+1 space-time. Here the non-vanishing contribution comes from indices running in the space part only. By contracting the indices we can obtain a scalar function which meaning might be clearer
\begin{eqnarray}
\frac{\delta H}{\delta g }=g^{\mu \nu } \frac{\delta H}{\delta g^{\mu \nu} } = - \frac{\hbar^2}{2m}  \left (    \partial^{\mu} \partial_{\mu} -   \hat{\Gamma}^{\nu}_{\mu \mu} \partial_{\nu}   \right) .
\end{eqnarray}
Indeed, for almost flat space $g_{\mu \nu} = \eta_{\mu}^{\nu} + h_{\mu \nu} $ and keeping the zero-th order terms
\begin{eqnarray}\label{eq: delta H / delta g}
\frac{\delta H}{\delta g } \approx - \frac{\hbar^2}{2m}  \partial^{\mu} \partial_{\mu} = \frac{p^2}{2m}=H_{\textrm{flat} }.
\end{eqnarray}
This equals the energy operator for the quantum gas in flat space-time. The variation of the energy operator with respect to the components of the affine connection is given by
$\delta H = - \frac{\hbar^2}{2m}    \left (     - \hat{g}^{jk}  \delta \hat{\Gamma}^{l}_{jk} \partial_{l}   \right) 
$ which yields
\begin{eqnarray}
\frac{\delta H}{\delta \Gamma^{\alpha}_{\mu \nu}} = - \frac{\hbar^2}{2m}   \left (     - \hat{g}^{\mu \nu} \partial_{\alpha}   \right)
\end{eqnarray}
an expression which for almost flat space reduces to
\begin{eqnarray}\label{eq:delta H/ delta Gamma}
\frac{\delta H}{\delta \Gamma^{\alpha}_{\mu \nu}} \approx \frac{i \hbar}{2m}   \left (     - i \hbar \partial_{\alpha}   \right) = \frac{i \hbar}{2m}  p_{\alpha} .
\end{eqnarray}
This is proportional to a component of the momentum defined for the flat space case.

The material part of the theory might be re-defined with the use of another expression for the kinetic energy term:
\begin{eqnarray}\label{eq:LB_g only}
H= - \frac{\hbar^2}{2m} \Delta_{LB}= - \frac{\hbar^2}{2m} \frac{1}{\sqrt{|\hat{g}|}} \partial_{j} \left( \sqrt{|\hat{g}|} \hat{g}^{jk} \partial_{k}  \right), 
\end{eqnarray}
where $\Delta_{LB}$ is the Laplace-Beltrami operator defined on a three-dimensional hyperplane of space-time. $|\hat{g}|=\mathrm{det} (\hat{g})$ is the determinant of the three-dimensional metric.  With this definition for the energy of the condensate 
\begin{eqnarray}\label{eq: delta H/ delta Gamma =0}
\frac{\delta H}{\delta \Gamma^{\alpha}_{\mu \nu}} = 0
\end{eqnarray}
the variation with respect to the affine connection is vanishing and the major difficulty in solving the theory can be transferred to the metric part. Thus the expression (\ref{eq: delta Theory / delta Gamma}) can be reduced to (\ref{eq:vollick}) since the entropy of the material part, that is the entropy of the quantum system is a function of its energy:
\begin{eqnarray}
S_{QM}(H)=-\mathrm{tr}\left( \rho \log \rho \right) =\frac{1}{Z} \mathrm{tr} \left( \beta H e^{-\beta H} \right) + \log Z,
\end{eqnarray}
where $Z=\mathrm{tr} \; e^{-\beta H}$ is the statistical sum emerging from the condition $\mathrm{tr} \; \rho = 1.$ Here the statistical operator is equal to (see Appendix)
\begin{eqnarray}\label{eq:rho(H)}
\rho=\frac{e^{-\beta H}}{Z},
\end{eqnarray}
where $\beta=1/k_B T.$ Here $k_B$ is Boltzmann's constant and $T$ the absolute temperature.

Now we are in a position to write the variation of the entropy of the quantum system in terms of the variation of its total energy operator:
\begin{eqnarray}\label{eq:delta S}
\delta S_{QM} &=&  \frac{S_{QM} -\log Z}{Z} \mathrm{tr} \left(  \beta \delta H e^{-\beta H}  \right) - \frac{1}{Z}\mathrm{tr} \left( \beta H \beta \delta H e^{-\beta H}  \right).
\end{eqnarray}

The value of this expression is revealed if one considers (\ref{eq:LB_g only}), therefore (\ref{eq: delta H/ delta Gamma =0}) and the variation of the entropy with respect to the components of the affine connection is vanishing. The final entropic field equation is then
\begin{eqnarray}\label{eq: final entropic Gamma}
\frac{\delta \sqrt{-g}  S_G}{\delta \Gamma^{\alpha}_{\mu \nu}}=-\frac{\delta \sqrt{-g}  S_{QM}}{\delta \Gamma^{\alpha}_{\mu \nu}}=0,
\end{eqnarray}
which solution is given by (\ref{eq:connection}) and (\ref{eq: R}).

Regarding the first form of the Hamiltonian (\ref{eq:LB g&Gamma}),  the variation of the entropy with respect to the components of the metric tensor becomes
\begin{eqnarray}\label{eq:delta S/delta g}
\nonumber \frac{\delta S_{QM}}{\delta g_{\mu \nu}} &=&  \frac{S_{QM} -\log Z}{Z} \sum_a   \beta   
\left\langle \psi_a \right|  \frac{\delta H}{\delta g_{\mu \nu}} \left| \psi_a \right\rangle 
e^{-\beta E_a}  \\
 &&- \frac{1}{Z}\sum_b   \beta E_b \beta 
\left\langle \psi_b \right|  \frac{\delta H}{\delta g_{\mu \nu}} \left| \psi_b \right\rangle e^{-\beta E_b} 
\end{eqnarray}
where $E_a= \left\langle \psi_a \right| H \left| \psi_a \right\rangle .$ Therefore, the contracted quantity for almost flat space $H \approx H_{\textrm{flat} },$ that is using (\ref{eq: delta H / delta g}),
\begin{eqnarray}
\frac{\delta S_{QM}}{\delta g} \approx \left( S_{QM} -\log Z \right)^2  
- \beta^2  \langle E^2 \rangle, 
\end{eqnarray}
where $\langle E^2 \rangle$ denotes mean squared energy. The geometric field part (\ref{eq: R}) is given by
\begin{eqnarray}\label{eq:R(S,logZ)}
R \approx \frac{3 L^{-2}}{ -2 S_{QM} + \left( S_{QM} - \log Z \right)^2  
- \beta^2  \langle E^2 \rangle   },
\end{eqnarray}
which also solves (\ref{eq:connection}).

Next, the variation with respect to the affine connection gives
\begin{eqnarray}\label{eq:delta S/delta Gamma}
\nonumber \frac{\delta S_{QM}}{\delta \Gamma^{\alpha}_{\mu \nu}} &=& \frac{S_{QM} -\log Z}{Z} \sum_a   \beta 
\left\langle \psi_a \right|  \frac{\delta H}{\delta \Gamma^{\alpha}_{\mu \nu}}  \left| \psi_a \right\rangle 
e^{-\beta E_a}  \\
&&- \frac{1}{Z}\sum_b   \beta E_b \beta 
\left\langle \psi_b \right|  \frac{\delta H}{\delta \Gamma^{\alpha}_{\mu \nu}}  \left| \psi_b \right\rangle e^{-\beta E_b} 
\end{eqnarray}
which due to (\ref{eq:delta H/ delta Gamma}) becomes 
\begin{eqnarray}
\frac{\delta S_{QM}}{\delta \Gamma^{\alpha}_{\mu \nu}}  \approx \frac{ i \hbar}{2m}  \beta (S_{QM} -\log Z) \,  \langle p_{\alpha} \rangle - \frac{ i \hbar}{2m}  \beta^2 \langle E \, p_{\alpha} \rangle, &&
\end{eqnarray}
where $\langle p_{\alpha} \rangle$ denotes the mean value of the $\alpha$ component of the momentum and $\langle E \, p_{\alpha} \rangle$ is the mean value of combined operator $Hp_{\alpha}.$ Note, for a quantum condensate this quantity is vanishing due to the vanishing of the canonical momentum
\begin{eqnarray}
\nonumber \frac{\delta S_{QM}}{\delta \Gamma^{\alpha}_{\mu \nu}} &\approx&0.
\end{eqnarray}
and for a quantum condensate (\ref{eq: final entropic Gamma}) still applies.
In this case, the equation for the connection (\ref{eq:connection}) holds as long as $f'(R)$ can be expressed as a function of the metric components and its derivatives. This requirement is fulfilled by the expression for $R,$ that is (\ref{eq:R(S,logZ)}).

Let us try to estimate the large volume $V \to \infty$ limit to the statistical sums $S_{QM}$ and ${\delta S_{QM}}/{\delta g}$ using the standard statistical mechanics rule\cite{Huang}
\begin{eqnarray}
\nonumber  E_a \to \frac{p^2}{2m} \quad \&\quad \sum_a \to \frac{V}{h^3}\int_0^{\infty} dp \; 4\pi p^2.
\end{eqnarray}
Effectively,
\begin{eqnarray}
\nonumber \frac{\delta S_{QM}}{\delta g} \to - \frac32,  &&
S_{QM} \to \frac32 + \log \frac{V}{\lambda_{th}^3},\\
- 2 S_{QM}  +  \frac{\delta S_{QM}}{\delta g}&\to& - \frac92 - 2\log \frac{V}{\lambda_{th}^3},
\end{eqnarray}
where $\lambda_{th}=\sqrt{ \frac{2 \pi \hbar^2 }{m k_B T} } $
is the thermal wavelength, which for the purpose of building an intuition takes the following characteristic values: i.) hydrogen atom at the temperature of the Universe 2.73K $\lambda_{th}\approx 1 \textrm{nm}$; ii.) liquid helium at lambda point 2.17K $\lambda_{th}\approx .6 \textrm{nm}$; iii.) superconducting Cooper pairs at liquid helium temperature 4.2K $\lambda_{th}\approx 26 \textrm{nm}.$

As a result the  entropic field equations have the following behaviour at large volume
\begin{eqnarray}
R\approx  - \frac{1}{ L^2 \left(\frac23  \log \frac{V}{\lambda_{th}^3} \right) }
\end{eqnarray}
which entails cosmological consequences as the large volume limit suggests that if the principle matter in the universe is hydrogen gas and the volume of the present epoch Universe is $V\approx 10^{80} \textrm{m}^3,$ \cite{sizeU} then $\frac23 \log \frac{V}{\lambda_{th}^3} \propto \times10^2.$ Therefore, the length-scale constant in the expression for the entropy density of the gravitational (geometric) field is $L=\sqrt{\frac{1}{ 10^2 \Lambda}  },$ where $\Lambda$ is the cosmological term, estimated at $10^{-52} \textrm{m}^{-2},$ that is $L$ is of the order of one tenth of the Einstein length, then
\begin{eqnarray} \label{eq:R Lambda}
R\approx  - \Lambda.
\end{eqnarray}
Therefore, {\it  the quantum matter content can account for the small negative curvature of space-time, which we term dark-energy.}

Now, suppose the quantum condensate is a Bose-Einstein gas at ground state, that is $S_{QM}=0$ and
\begin{eqnarray}
\langle E^2 \rangle= \frac{\left( \log Z \right)^2}{\beta^2}= \left( 2 \log Z \right)^2 \left( \frac12 k_B T \right)^2
\end{eqnarray}
the entropic field equation (\ref{eq:R(S,logZ)}) predicts
\begin{eqnarray}
R \to \infty,
\end{eqnarray}
that is the geometry of space time diverges while the entropy of the geometric field $S_G \to 0.$ For a non-interacting Bose-Einstein gas the conditions for this phenomenon can be further specified 
\begin{eqnarray}
 2 \log Z = 3 \quad \lambda_{th} = {V^{1/3} }/{ \sqrt{e} }.
\end{eqnarray}
The general solution for $S_{QM} \neq 0$ leading to singularity is given by
\begin{eqnarray}
\log Z =S_{QM} + \sqrt{2S_{QM} +  \beta^2 \langle E^2 \rangle }.
\end{eqnarray}
Suppose, we have a Cooper pair gas in the interior of a superconductor at liquid helium temperature. The thermal wavelength in this case is of the order of 26 nm. Since $V^{1/3}$ is a characteristic length for the gas, we may assume it is few times the coherence length for the Cooper pairs $ \xi_0 \approx 80 \textrm{nm} $ in the case of Pb ( $ V^{1/3} /\sqrt{e} \approx 48 \textrm{nm}$). At temperature of around 1mK, the Cooper pair thermal wavelength is $\lambda_{th} \approx 1.7 \mu\textrm{m},$ which is appropriate to test the predicted effect.

The predicted divergence in the underlying geometry as the quantum gas falls into coherent state with particular characteristics seems to violate energy conservation and appears thermodynamically improbable as $S_G + S_{QM} \to 0,$ which is not a preferred process. However, {\it if such a process is possible locally, it represents a verifiable prediction of the theory.} The curvature of space-time need not diverge into singularity, but at this state any energetic input in the quantum system can channel into the geometric field. Similar process with an underlying energy conservation law emerges in condensed matter context\cite{Victor}.

\section{Energetic aspect}

With regard to the energetic aspect of the entropic theory, it can either be i.) associated with the temperature of the compound system in a way similar to black hole thermodynamics
\begin{eqnarray}
d E= T dS \propto \kappa \; d \textrm{Horizon area},
\end{eqnarray}
where $\kappa$ is the surface gravity on the horizon, provided we have an expression for the temperature $T,$ or ii.)  define the energy of the geometric field (the energy of the matter fields is given by their Lagrangian) and then obtain an expression for the temperature if neccessary. 

The energy of the geometric field, by virtue of the Einstein theory of gravity, has been a controversial issue since the inception of the theory principal reason being that the canonical energy-momentum pseudotensor (Landau-Lifshitz pseudotensor) is made up of first derivatives of the metric and these vanish where the frame is locally inertial at any chosen point, that is it doesn't contain much useful information \cite{Landau}. The Einstein pseudotensor \cite{Einstein} fails to be symmetric. The Cooperstock's hypothesis\cite{Cooper}, namely the geometric energy only exists where the energy-momentum tensor is non-vanishing fails to account for the existence of gravitational waves\cite{LIGO}.

Here we will use an unique definition of the local energy of the gravitational field
which stems from the framework of\cite{Victor}. A quantum mechanical system 
constrained to abide a curved hyper-plane of space-time, produces a geometric potential from the kinetic term. An energy conservation relation in a material  medium involving the geometric field can be demonstrated. The energy of the geometric field is given by
\begin{equation}
\mathcal{E}_G=\frac{\hbar^2}{24m}  R^{3D} 
\end{equation}
where $R^{3D}$ is the induced three dimensional Ricci scalar curvature, and $m$ is the mass of the bosons subjected to the field (in the framework of \cite{Victor} these are the Cooper pairs). It is related to the four-dimensional Ricci curvature of space-time with the relation $R=\frac{4}{3} R^{3D}$ \cite{Ficken}.

We are now in a position to state a definition (up to numerical factor) of the local energy of the gravitational field
\begin{equation}
\mathcal{E}_G=\frac{\hbar^2}{m^{*}}  R. 
\end{equation}
Here $m^{*}$ is an unknown parameter in the theory with the dimension of mass (Planck mass is an option $m^{*}=m_p=\sqrt{\hbar c / G}$). The above expression can serve as an energy density to the energy functional
\begin{equation}
E_G=\frac{\hbar^2}{m^{*}}  \int  R \sqrt{-g} d^4 x
\end{equation}
which can be extended and include matter field via their Lagrangian density $\mathcal{L}_{M}:$
\begin{equation}
  \int \left( \frac{\hbar^2}{m^{*}} R  + \mathcal{L}_{M} \right) \sqrt{-g} d^4 x = \textrm{const.}
\end{equation}

The similarity with the Einstein-Hilbert action is obvious and therefore its variation would yield the same equations of motion. In the case of matter free Universe, that is the ''vacuum'' case we have:
\begin{eqnarray}
\delta E_G= 0 &\Rightarrow& R^{\mu \nu} - \frac12 g^{\mu \nu}R =0.
\end{eqnarray}
The entire repertoire of classical General Relativity is preserved. The entropic aspect implies $\lim_{R \to 0}{S_G}=\infty$ which is thermodynamically favourable up to the non-existence of affine connection (\ref{eq:connection}) in an empty Universe, in accord with the Mach's principle of relativity. However, a minimal amount of matter would lead to the restoration of the affine connection aspect of the geometric field. 

One last issue deserves attention, namely the quantum aspect. Since information content can be represented as a set of discrete units called bits, it is only natural to identify the bit as the quantum of information. The amount of dimensionless entropy corresponding to one bit is $\log2,$ therefore a quantum of curvature can be defined with the help of (\ref{eq:S_G}) and (\ref{eq:R Lambda}):
\begin{eqnarray}
R=\frac{1}{L^2 \log2} \sim \frac{\Lambda}{\log2},
\end{eqnarray}
where $\Lambda$ is the cosmological constant. The quantum of geometrical energy upon the use of Planck's mass becomes
\begin{equation}
\mathcal{E}_G=\hbar^{3/2}  \sqrt{ \frac{G}{c}}  \frac{\Lambda}{\log2} \sim 10^{-93} \; \textrm{eV}, 
\end{equation}
which is vanishing but non-zero. If this is the quantum of energy of gravitational waves, that is the graviton's mass $m_g,$ it is well within the bound $m_g < 10^{-22}$ eV/c$^2$ \cite{LIGO}.

\section{Conclusions}

In conclusion, we would like to go through the main ideas explored in the paper. We derive entropic equations for gravity from a Palatini variational principle based on conservation of information (maximalisation of entropy). Palatini variation is used in order to reduce the order of the governing equations and we reckon is not essential to the main results, which are: i.) a local invariant expression for the entropy of the geometric field can be defined; ii.) the geometric field does not exist in an empty of matter Universe, that is the geometric field can be constructed in accord with the Mach's principle; iii.) material entropy is geometry (lengthscale) dependent; iv.) matter can exchange information (entropy) with the geometric field; v.) a  quantum condensate can channel energy into the geometric field at a particular coherent state; vi.) an expression for the local energy density of the geometric field can be defined; vii.) Einstein vacuum equations emerge as the energy conservation aspect of the theory and are thermodynamically favoured provided some matter occupies space and viii.) the cosmological constant (the energy density of dark matter) emerges within the large volume limit in the theory pointing to the {\it non-existence of dark energy in an empty universe.} 

The author wishes to acknowledge inspiring discussions with Hristo Dimov, Dimitar Marvakov and Rossen Dandoloff.

\section*{Appendix}

Here the statistical operator is equal to
\begin{eqnarray}
\rho=\frac{e^{-\beta H}}{Z},
\end{eqnarray}
where $\beta=1/k_B T.$ Here $k_B$ is Boltzmann's constant and $T$ the absolute temperature. $Z=\mathrm{tr} \; e^{-\beta H}$ is the statistical sum emerging from the condition $\mathrm{tr} \; \rho = 1.$

This solution emerges as a result of imposing a stationary state condition $\left[ H, \rho \right]=0$, that is the quantum system is in a pure state:
$\left[   H + \delta H  ,\rho + \delta \rho \right]=0,
$ which leads to
\begin{eqnarray}\label{eq: [H, delta rho] }
\left[   H ,\delta \rho \right]=-\left[  \delta H , \rho \right].
\end{eqnarray}
Inserting (\ref{eq:rho(H)}) in the last equation we get
\begin{eqnarray}
\nonumber \frac{1}{Z} \left[   H , - \beta \delta H e^{-\beta H} \right] - \frac{\delta Z}{Z^2} \left[   H , e^{-\beta H} \right] =- \frac{1}{Z} \left[  \delta H , e^{-\beta H} \right],
\end{eqnarray}
which can further be reduced using the relation $ \left[  A  , BC \right]= \left[  A  , B \right]  C + B \left[  A  , C \right]$ to
\begin{eqnarray}
- \beta  \left[   H , \delta H  \right]  e^{-\beta H} - \beta \delta H  \left[   H , e^{-\beta H} \right]  = -\left[  \delta H , e^{-\beta H} \right].
\end{eqnarray}
This last relation is fulfilled identically provided
$\left[   H , \delta H  \right]  = 0 ,
$ that is if the Hamiltonian commutes with its variation (\ref{eq:rho(H)}) is a solution to (\ref{eq: [H, delta rho] }) and the system remains in a pure state: $\left[ H, \rho \right]=0.$ Explicitly,
$\delta \rho=\delta \; {e^{-\beta H}}/{Z}.$

\end{document}